\documentclass[accepted]{article}

\usepackage{microtype}
\usepackage{graphicx}
\usepackage{subfigure}
\usepackage{booktabs} % for professional tables

\usepackage{multirow}
\usepackage[accepted]{icml2024}

\usepackage{amsmath}
\usepackage{amssymb}
\usepackage{mathtools}
\usepackage{amsthm}

\usepackage[capitalize,noabbrev]{cleveref}

\theoremstyle{plain}

\theoremstyle{definition}

\theoremstyle{remark}

\newtheorem*{problem*}{Problem}
\newtheorem{problem}{Problem}
\newcommand{\rom}[1]{\uppercase\expandafter{\romannumeral #1\relax}}
\newcommand{\GND}{\mbox{$\mathit{GND}$}}
\newcommand{\VOUT}{\mbox{$\mathit{VOUT}$}}
\newcommand{\VIN}{\mbox{$\mathit{VIN}$}}
\newcommand{\Sb}{\mbox{$\mathit{Sb}$}}
\newcommand{\Sa}{\mbox{$\mathit{Sa}$}}

\icmltitlerunning{LaMAGIC: Language-Model-based Topology Generation for Analog Integrated Circuits}

\begin{document}

\twocolumn[
\icmltitle{
%Automated Analog Circuit Design via \\ Language Model-based Topology Generation}
%LaMAGIC: Language-Model-based Topology Generation for Analog Integrated Circuit}
LaMAGIC: \underline{La}nguage-\underline{M}odel-b\underline{a}sed Topology \underline{G}eneration \\for Analog \underline{I}ntegrated \underline{C}ircuits}



\icmlsetsymbol{equal}{*}

\begin{icmlauthorlist}
\icmlauthor{Chen-Chia Chang}{ibm,duke}
\icmlauthor{Yikang Shen}{mit}
\icmlauthor{Shaoze Fan}{nj}
\icmlauthor{Jing Li}{nj}
\icmlauthor{Shun Zhang}{mit}
\icmlauthor{Ningyuan Cao}{und} \\
\icmlauthor{Yiran Chen}{duke}
\icmlauthor{Xin Zhang}{ibm,mit}
\end{icmlauthorlist}

\icmlaffiliation{ibm}{IBM T. J. Watson Research Center}
\icmlaffiliation{duke}{Duke University}
\icmlaffiliation{mit}{MIT-IBM Watson AI Lab}
\icmlaffiliation{nj}{New Jersey Institute of Technology}
\icmlaffiliation{und}{University of Notre Dame}

\icmlcorrespondingauthor{Chen-Chia Chang}{chenchia.chang@duke.edu}
\icmlcorrespondingauthor{Xin Zhang}{xzhang@us.ibm.com}

\icmlkeywords{Electronic design automation, Large language model, Analog design automation, Machine learning}

\vskip 0.3in
]



\printAffiliationsAndNotice{}  % leave blank if no need to mention equal contribution

\begin{abstract}
% The manual design of analog circuit is a highly specialized task because of its high complexity and the required expertise.
% It is a time-consuming process that often requires iterative circuit refinement to achieve desired circuit performance, posing significant challenges to circuit designers.
% Inspired by this, we present a novel automated method that develops a generative language model targeting on power converter circuits, an application that requires designers to design different circuit topologies to meet various specifications of voltage conversion ratio and efficiency. 
% By feeding user-defined performance targets into our language model, it intelligently generates the circuit topology that has performance closed to the requirement.
% This approach leverages the generative strengths of language models to significantly accelerate the design process. 
% Experimental results show that our model can achieve around 95\% of success rate in generating circuit within 3\% of error for the target performance.

In the realm of electronic and electrical engineering, automation of analog circuit is increasingly vital given the complexity and customized requirements of modern applications. 
However, existing methods only develop search-based algorithms that require many simulation iterations to design a custom circuit topology, which is usually a time-consuming process.
To this end, we introduce LaMAGIC, a pioneering language model-based topology generation model that leverages supervised finetuning for automated analog circuit design.
LaMAGIC can efficiently generate an optimized circuit design from the custom specification in a single pass.
Our approach involves a meticulous development and analysis of various input and output formulations for circuit.
These formulations can ensure canonical representations of circuits and align with the autoregressive nature of LMs to effectively addressing the challenges of representing analog circuits as graphs. 
The experimental results show that LaMAGIC achieves a success rate of up to 96\% under a strict tolerance of 0.01. 
We also examine the scalability and adaptability of LaMAGIC, specifically testing its performance on more complex circuits. 
Our findings reveal the enhanced effectiveness of our adjacency matrix-based circuit formulation with floating-point input, suggesting its suitability for handling intricate circuit designs.
This research not only demonstrates the potential of language models in graph generation, but also builds a foundational framework for future explorations in automated analog circuit design.

% In addition, we develop several novel circuit formulations for LM to ensure canonical circuit representations that align with the autoregressive nature of LMs.
% A key innovation of La-MAGIC is to extend capabilities of LMs to graph generation.
% This advancement not only accelerates the design process but also offers a scalable solution to meet the evolving demands of this field.
% In the experiments, we thoroughly analyze 

% a novel approach employing supervised fine-tuning for language models (LMs) for efficient and accurate design of analog circuits. La-MAGIC diverges from traditional methods by implementing bespoke input and output formats that effectively represent graph-based circuit topologies, thereby ensuring both functional relevance and syntactic accuracy. Our exploration includes a thorough analysis of various data representations and the optimization of output formats, particularly focusing on the precision of defining essential parameters like duty cycle. A key innovation of La-MAGIC is the extension of LM's capabilities to encompass graph generation, enhancing the model's ability to efficiently generate optimized circuit topologies. This advancement not only streamlines the design process but also represents a significant leap forward in automated analog circuit design, offering a scalable and precise solution to meet the evolving demands of this field.

\end{abstract}

\section{Introduction}

% \red{1. Difficulty of analog circuit design. Need of automatic circuit design}

\begin{figure}[t]
\centering
\includegraphics[width=1\linewidth]{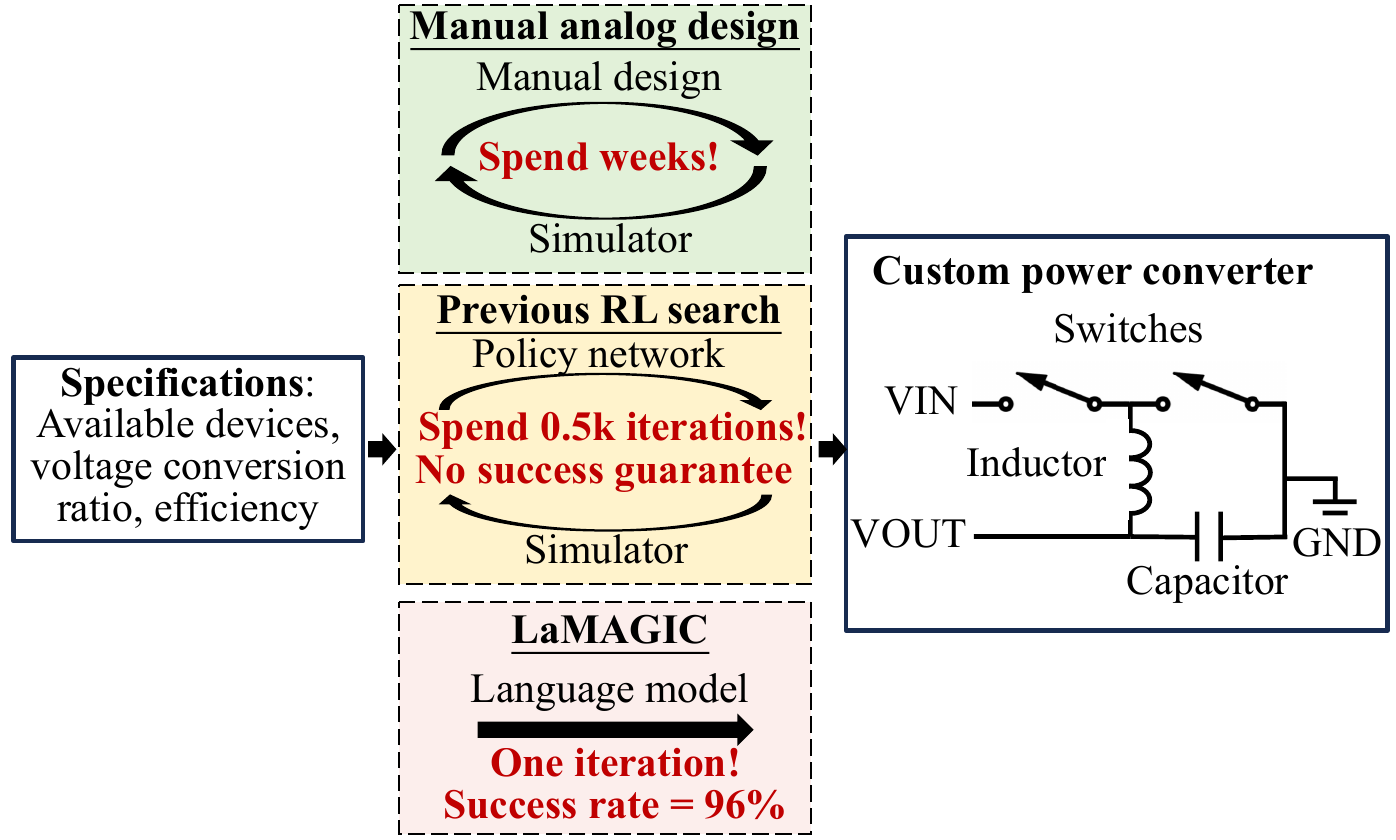}
\vspace{-2mm}
\caption{
The scenario comparison between manual analog circuit design, previous RL search method~\cite{fan2021specification}, and our language model LaMAGIC.
}
\label{fig:manual vs automation}
% \vspace{-5mm}
\end{figure}

Analog circuit design, encompassing a broad range of applications, poses significant challenges due to its inherent complexity of the schematic and the precision of performance required in its execution. 
This complexity is particularly pronounced in the realm of power converters, which have become ubiquitous in an array of electronic and electrical devices. 
With the advent of diverse and customized electrical systems, such as electric vehicles, self-powered Internet of Things devices, and wearable or implantable biosensors, the demand for custom-designed power converters to meet specific supply power standards has surged. 
These converters, each with unique design specifications including voltage conversion ratio and power efficiency, exemplify the intricate and varied nature of analog circuit design. 
Traditional design methodologies, largely depending on pre-existing circuit topologies and extensive manual optimization, are increasingly inadequate in addressing the increasing needs of these applications. 
This reliance on conventional approaches not only prolongs the design process but also limits the potential for novel solutions in rapidly evolving domains. 
This gap highlights a critical need for an automated circuit design framework, capable of efficiently generating and optimizing high-quality power converter topologies based on specific design specifications. 
% Such a framework can not only accelerate the design process for these essential components but also pave the way for broader advancements in the field of analog circuit design.

% \red{2. Previous work for automatic circuit design}

While there have been notable efforts to address the challenges of automated analog circuit design, as highlighted in recent studies~\cite{fan2021specification, zhao2022analog, lu2023automatic}, these initiatives are yet to fully overcome the inherent complexities of the field. 
The approach~\cite{fan2021specification} leverages a reinforcement learning (RL)-based tree sampling process for automating power converter design. 
However, this method exhibits limitations in terms of scalability and practical applicability, especially when generating circuits with varying performance specifications. 
It requires approximately 500 simulation queries for convergence each time a new circuit design is initiated, underscoring a significant challenge in efficiently adapting to different specifications. 
The other methods~\cite{zhao2022analog, lu2023automatic} also develop search-based algorithm to sample promising circuit, which requires lots of simulations when querying a new specification.
This bottleneck highlights the necessity for a more direct and efficient generation method that can achieve desired performance criteria in a single iteration. 
Such a method can not only accelerate the automated design process but also enhance the practicality and applicability of automated analog circuit design, especially in rapidly evolving and diverse application areas.

% \red{3. LLM and why LLM. Why this task is suitable for LLM and vice versa IMPORTANT!!!. Extend LLM to support graph generation.}

The advent of large language models (LLMs)~\cite{radford2019language, raffel2020exploring, chung2022scaling, nijkamp2023codegen2} has opened new frontiers in numerous fields, demonstrating remarkable capabilities in understanding and generating complex patterns and structures. 
This makes LLMs particularly promising for tackling the intricate challenges of automated analog circuit design. 
The core strength of LLMs lies in their ability to process and synthesize vast amounts of data, learning underlying patterns and relationships. 
In the context of analog circuit design, this translates to the potential for LLMs to understand and generate the nuanced and often non-linear relationships between different circuit components and their performance characteristics. 
% Also, the sequence-to-sequence nature of Language Models (LMs) equips them with the capability to generate analog circuits by sequentially predicting components and their connections, effectively translating design specifications into complete circuit topologies.
% Moreover, LLMs' proficiency in natural language processing can be leveraged to interpret and translate design specifications into functional circuit designs, bridging the gap between high-level requirements and detailed technical implementation. 

However, to fully harness this potential in the field of analog circuit design, a significant extension of LLM capabilities is required, specifically in the domain of graph generation. 
Analog circuits can be effectively represented as graphs, where components are nodes and connections are edges, encapsulating both the structural and functional aspects of the design. 
Extending LLMs to support graph generation would enable them to directly generate circuit topologies from specifications with high efficiency and accuracy.
% considering not just the individual components but also their interconnections and overall design architecture. 
% This extension would mark a transformative step in the field, allowing for the direct generation of optimized, application-specific circuit designs with high efficiency and accuracy.

% \red{4. method}

This paper proposes LaMAGIC, an LM-based topology generation model for automated analog circuit design, especially for power converter applications. 
The application scenario is shown in Figure~\ref{fig:manual vs automation}.
To the best of our knowledge, we are the first to solve this problem through generative modeling and LM methodology.
We develop a novel approach centered around the supervised finetuning (SFT) to customize the LM into the domain of intricate analog circuit design. 
The core of LaMAGIC is the meticulous examination of data formats and their influence on the model's capabilities. 
We propose and investigate various innovative input and output formulation for circuit generation to synchronize with the operational dynamics of LMs and effectively capture graph-based circuit representations using the autoregressive loss function of LMs.
With the sequence-to-sequence modeling in LM, we can abstract the circuit generation into a process of sequential component or connection selection by utilizing the preceding subgraph information.
In addition, we explore the effects of representing circuit specifications with different data types i.e., floating-point numbers versus characters. 
This aims to explore the effectiveness of feeding numerical input into LM to help the circuit learning.

Our contributions can be summerized as follows:
\vspace{-3mm}
\begin{itemize}
    \item  
    We introduce LaMAGIC, a pioneering approach that adapts LMs to the domain of analog circuit design through SFT. This enables the efficient one-shot generation of custom circuit designs from user-defined specifications.
    
    \item  We propose multiple novel circuit formulations designed to enhance circuit generation by (1) ensuring canonical representations, (2) enhancing compatibility with the autoregressive training methods and loss functions of the LM, and (3) utilizing float inputs to optimize LM's processing capabilities.
    \item Experimental results show that our model achieves superior 0.96 success rate under a strict tolerance of 0.01. 
    We further conduct an extensive evaluation of the model’s performance and its adaptability to more complex circuit designs. 
    These evaluations provide critical insights into the effectiveness of different formulations, contributing significantly to future advancements in this field.
    \item  By extending the functionalities of LMs to include graph generation, LaMAGIC marks a significant step forward in the generation of optimized circuit topologies directly from specifications. This expansion has the potential to inspire similar applications in other areas of graph generation within the LM domain.

\end{itemize}
These contributions collectively represent a significant advancement in the field of automated analog circuit design, particularly in improving the efficiency, accuracy, and applicability of LMs in generating custom and application-specific circuit designs.

\section{Preliminaries}

\subsection{Analog Circuit Design}
% \red{goal of analog circuit design}

In the rapidly evolving domain of electrical and electronic engineering, automated analog circuit design, particularly for custom power converter applications, is very importance. 
This automation process aims to produce customized power converters without human interference, adhering to specific design specifications. 
Key among these specifications are the voltage conversion ratio  and the power conversion efficiency. 
The voltage conversion ratio is defined as the ratio between input and output voltages, while the power conversion efficiency is the ratio of input power to output power. Another crucial aspect of power converter design is the duty cycle, which controls the duration of switches within the circuit, thereby influencing the output voltage and the overall performance of the circuit.

\begin{figure}[t]
\centering
\includegraphics[width=\linewidth]{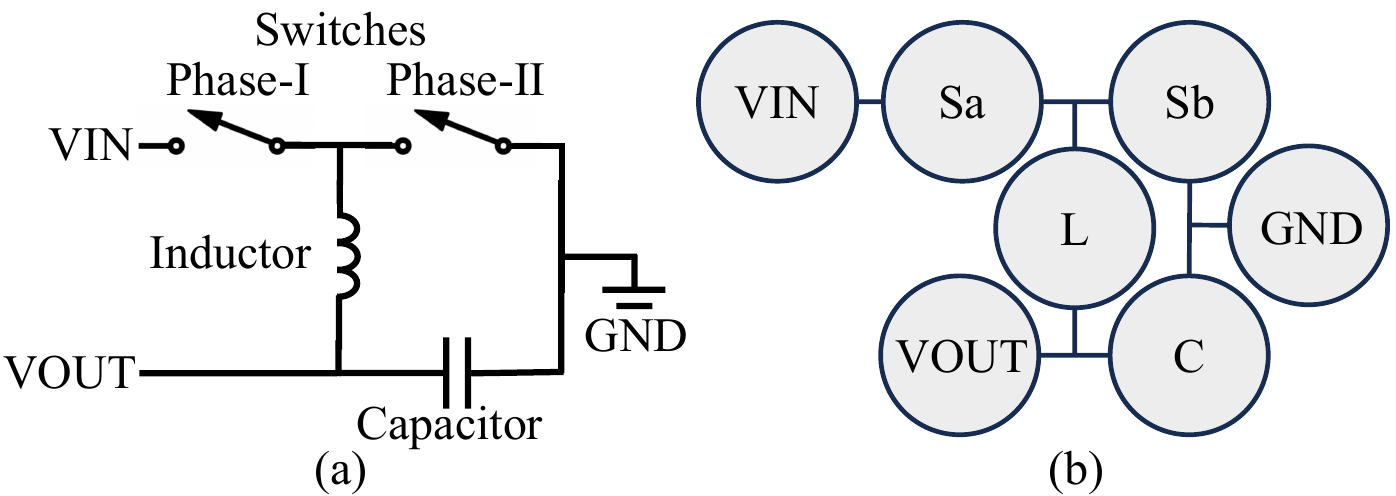}
\vspace{-3mm}
\caption{ (a) An example power converter circuit and (b) its corresponding hypergraph representation.
}
\vspace{-3mm}
\label{fig:example circuit}
\end{figure}

The circuit topology $G$ is conceptualized as a hypergraph consisting of vertices $V$ and hyperedges $E$. The vertices $V$ include various analog devices and three external terminal ports. 
The device (or called \textit{component} in later context), including capacitors $C$, inductors $L$, phase-I switches $\Sa$, and phase-II switches $\Sb$, is connected to other devices or ports via two outgoing edges. The terminal ports are the voltage input port $\VIN$, the voltage output port $\VOUT$, and the ground $\GND$, each with a single outgoing edge. The hyperedges $E$ symbolize the connections between these devices and ports.
An example of the power converter and its hypergraph representation is shown in Figure~\ref{fig:example circuit}.

% \subsection{Graph Generation}
% \red{previous work for graph generation}

\subsection{Language Model}
% \red{LM training}

LMs~\cite{radford2019language, raffel2020exploring, chung2022scaling, nijkamp2023codegen2}, especially those using autoregressive training, are pivotal in natural language processing. Autoregressive LMs learn to predict the next token in a sequence based on previous tokens, utilizing an autoregressive loss function. This function calculates the loss as the negative sum of log probabilities for each predicted token, given the preceding ones. Formally, for a sequence $x_1, x_2,\ldots,x_n$, the loss $\ell$ is:
$
\ell = - \sum_{i=1}^n \log P(x_i|x_1, x_2, ..., x_{i-1}).
$
This method trains LMs to capture complex sequential patterns, essential for generating contextually and syntactically coherent sequences. 
In automated analog circuit design, this approach is particularly beneficial. 
During the training process, LM can learn how to base on the previous subgraph information to decide the next component and connection.
% \red{The LM's capability to sequentially generate elements of a circuit, considering previously generated components, is analogous to constructing a well-formed narrative in language, making this loss function a promising choice for designing circuit topologies.}

\subsection{Problem Formulation}

% In addition to the fundamental design specifications of voltage conversion ratio and power conversion efficiency, the duty cycle is further explored in the context of design options. 
The objective of our model is to design circuit topologies and select appropriate duty cycles to achieve specified voltage conversion ratio and power conversion efficiency. Within our design framework, we consider a range of duty cycle options: \{0.1, 0.3, 0.5, 0.7, 0.9\}. These options provide a framework for varying the ON times of switches to meet specific performance criteria.
Based on these considerations, we define two distinct problem scenarios:

\begin{problem}[Edge Generation]
Given vertices $V$, a target voltage conversion ratio $r$, and an efficiency $\eta$, the task is to generate the connections $E$ and determine the duty cycle $s\in\{0.1, 0.3, 0.5, 0.7, 0.9\}$ to construct a circuit that satisfies both $r$ and $\eta$.
\end{problem}
\begin{problem}[Topology Generation]
In the scenario where device requirements are not predefined, the challenge expands to generating both vertices $V$ and their corresponding connections $E$, along with deciding the duty cycle $s$. The goal remains to construct a circuit meeting $r$ and $\eta$.
\end{problem}

\section{Power Converter Dataset Construction}

The foundation of an effective automated analog circuit design system lies in the diversity of the dataset upon which the model is trained. 
In our main experiment (Section~\ref{sec:345 comp experiment}),  we construct a dataset by randomly sampling topologies of 3, 4, and 5-component circuits. 
This range was chosen to encapsulate the varying degrees of complexity typical in power converter circuits, thereby ensuring that our model will be learned to handle a variety of design scenarios.
Additionally, we ensure that each topology is not isomorphic, which not only prevents redundancy in our dataset but also reinforces the diversity of circuit designs. 
For each random-sampled topology, we generate five circuits using different duty cycles \{0.1, 0.3, 0.5, 0.7, 0.9\} according to our design space.
Then, we simulate each circuit with NGSPICE~\cite{ngspice} to compute the corresponding voltage conversion ratio and efficiency. 
Next, we pruned out the invalid topologies reported by the simulator.
The final dataset comprises input features including the simulated voltage conversion ratio and efficiency, and the output consisting of the circuit topology and the duty cycle. 
For 3, 4, and 5-component circuits, we have 1k, 14k, and 117k different data points respectively, since the design space will exponentially grow up along with the component numbers.
In total, we randomly split around 120k data points for training and 12k for evaluation. 

In the subsequent experiment (Section~\ref{sec:6 comp experiment}), we extend our dataset to include 6-component circuits, aiming to assess the transferability of our model previously trained on 3, 4, 5-component circuits.
This step is crucial in evaluating the model's ability to generalize its learning and adapt to more complex circuit designs.
In total, we sample 76k different 6-component circuits and allocate 9k data points for evaluation.

% To construct a robust dataset for our power converter application, we sample random topologies comprising 3, 4, 5, and 6-component circuits.  
% This range was chosen to encapsulate the varying degrees of complexity typical in power converter circuits, thereby ensuring that our model would learn to handle a variety of design scenarios.
% Also, we ensure that each topology is not isomorphic, which not only prevents redundancy in our dataset but also reinforces the diversity of circuit designs. 

% For each randomly generated topology, we generate five circuits using different duty cycles \{0.1, 0.3, 0.5, 0.7, 0.9\} according to our design space.
% Then, we simulate each circuit with NGSPICE~\cite{ngspice} to compute the corresponding voltage conversion ratio and efficiency. 
% Next, we pruned out the invalid topologies reported by the simulator.
% The final dataset comprises input features including the simulated voltage conversion ratio and efficiency, and the output consisting of the circuit topology and the duty cycle. 
% In total, we randomly separated 120k data for training and 13k for evaluation. 
% \red{how many circuits you have for each component count?}

\begin{figure*}[t]
\centering
\includegraphics[width=\linewidth]{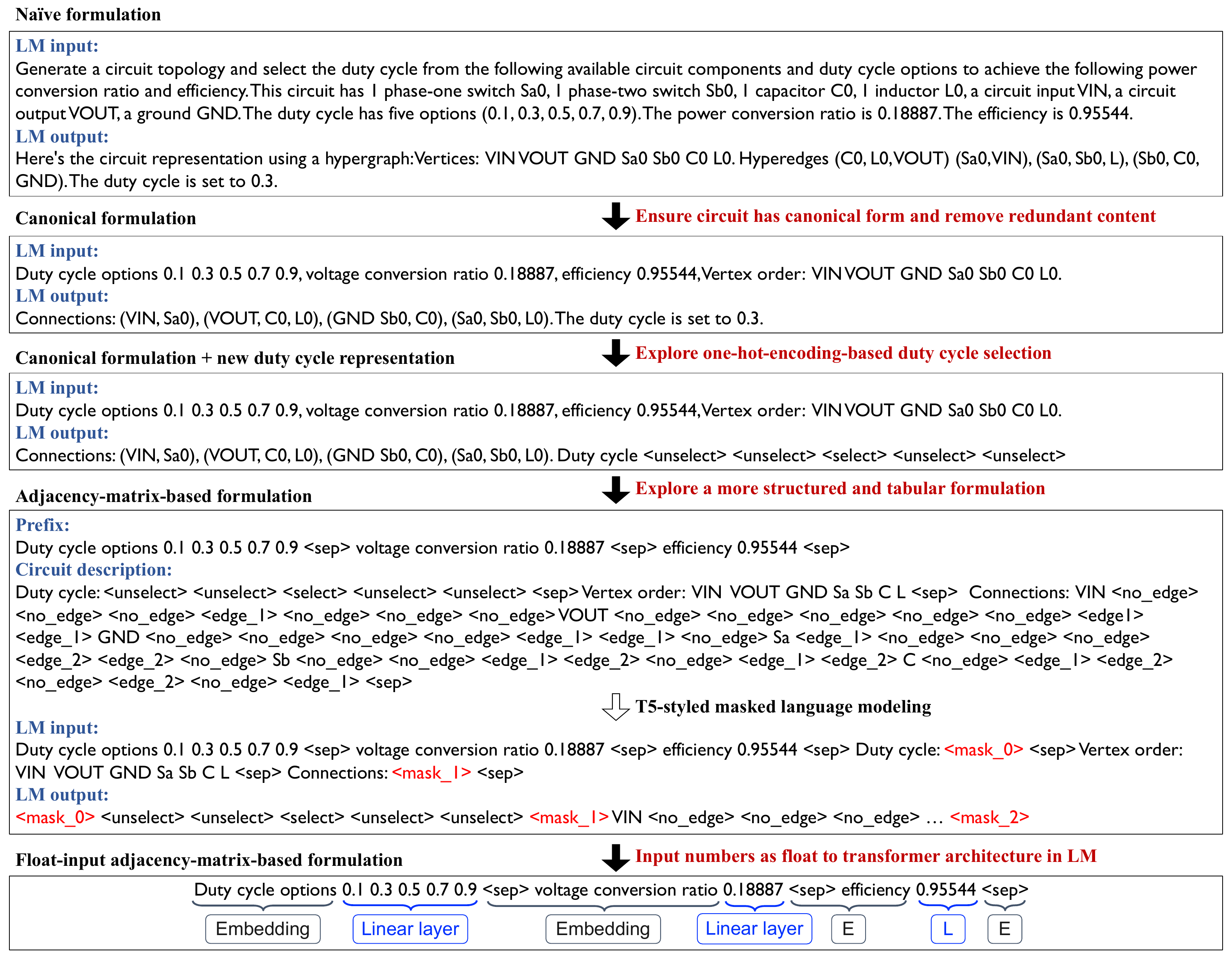}
\vspace{-5mm}
\caption{
A circuit example from Figure~\ref{fig:example circuit} representing by the na\"ive formulation, our canonical formulation, canonical formulation + new duty cycle representation, adjacency-matrix based formulation, and float-input adjacency based matrix formulation for edge generation task. 
In the top three formulation, we simply place the vertex-related description from LM input to output to achieve topology generation task. 
In adjacency-matrix based formulation, we employ T5-styled masked language modeling~\cite{raffel2020exploring} to perform circuit generation. 
For topology generation, we further mask the vertex order in LM input and place it to output.
% \red{AMBF is inconsistent with the next figure. Should only leave one. "inductors". inconsistent box sizes}
}
\label{fig:different formulation}
\end{figure*}

\section{Language Model-based Topology Generation} \label{sec:method}

% \red{Motivation using supervise fine tuning}
% \blue{figure to show prompt is not working}

% \red{Data format: Problem of naive format.
% In: Float
% Out: 1. Edge format cannot capture graph similarity under current loss. 2. Duty cycle representation 3. output order (graph first or duty circuit first 4. Not canonical form, e.g. two edge swap location result in diff lost)}

% \red{Intro to your data format}

% \red{training method: mask, model}
\subsection{Na\"ive Methods for Circuit Generation}

Using LLMs to revolutionize the labor-intense tasks is a promising trend given the generation ability and the natural language formats for input and output of LLMs.
Prompt engineering~\cite{liu2023pre} and SFT~\cite{nijkamp2022codegen} are the most famous approach to adapt LLMs to specialized domains.
However, the unique challenges of analog circuit design, which demand specialized expertise and interaction with simulators, present limitations for fundamental LLMs like GPT-4~\cite{lee2023benefits} when used prompt engineering. 
Our experiment with GPT-4, involving few-shot prompt engineering with 100 samples in the analog circuit design domain, revealed that the model failed to produce circuits meeting our specific performance requirements. 
This outcome illustrates the limitations of using general-purpose LLMs and prompt engineering for highly specialized tasks.

% However, given the uniqueness of this analog circuit design problem that requires strong expertise and interacts to simulators, fundamental LLMs like GPT4~\cite{lee2023benefits} cannot tackle with such specialized problems. 
% Thus, prompt engineering is not suitable for this tasks. \blue{(need the experiment to support this claim. may use few-shot learning for GPT-4)}
As a result, in this work, we aim to explore SFT to build an LM for analog circuit generation.
A na\"ive method is to formulate the problem into instruction-based context inspired by the recent success of instruction-based finetuning methods~\cite{chung2022scaling, alpaca}, as shown in the first formulation in Figure~\ref{fig:different formulation} for edge generation.
% \begin{itemize}
%     \item Input: Generate a circuit topology and select the duty cycle from the following available circuit components and duty cycle options to achieve the following power conversion ratio and efficiency. This circuit has 2 phase-one switches Sa0 and Sa1, 1 phase-two switch Sb0, 2 inductors L1 and L0, a circuit input VIN, a circuit output VOUT, a ground GND. The duty cycle has five options (0.1, 0.3, 0.5, 0.7, 0.9). The power conversion ratio is 0.18887. The efficiency is 0.95544.

%     \item Output: Here's the circuit representation using a hypergraph: Vertices: VIN, VOUT, GND, Sa0, Sa1, Sb0, L0, L1.
%     Hyperedges (GND, Sb0), (VIN, Sa0, L0), (L1, VOUT), (Sb0, Sa1, L0), (L1, Sa0, Sa1). The duty cycle is set to 0.1.
% \end{itemize}
This format offers an user-friendly approach, designing the input as a natural language instruction with detailed specifications.
% , and the output as a clear circuit representation using vertices, hyperedges, and duty cycle decisions. 
However, this seemingly straightforward formulation might present some challenges for LMs:
\begin{enumerate}
    \item Non-unique representations: The order of hyperedges can be permuted while still representing the same circuit structure, leading to multiple potential outputs representing for a single input. This non-uniqueness can complicate the learning process by causing ambiguity in the loss function calculation, potentially misguiding the model training.
    % \item Specialization vs. Language: The highly specialized nature of circuit design, replete with intricate relationships and rules, might not be fully captured by natural language instructions e.g., "Generate a circuit ...", potentially hindering the LM's comprehension.
    % \red{i don't think this make sense}
    \item Structured connection representation: A more structured format might be necessary for effectively representing connections, aligning better with the sequential nature of LM training and loss functions, to facilitate learning of connections and component selections.
    \item Precision of technical specifications: Presenting technical specifications such as voltage conversion ratio and efficiency in textual format might reduce precision, requiring additional effort from the LM to interpret these values correctly.
\end{enumerate}

% (2) 
% (3) Structured Connection Representation: A more structured format might be necessary for effectively representing connections, aligning better with the sequential nature of LM training and loss functions, thereby facilitating the learning of likely connections or component selections.

% (1) The order of hyperedges can be permuted to represent the same circuit structure, so there is no unique circuit representation given each input. 
% This will cause one-to-many mapping between input and label and will make learning harder.
% (2) The natural language instruction e.g., "Generate a circuit ..." may not help LM to interpret circuit because circuit design is a highly specialized domain that often involves complex relationships and rules that may not be captured by natural language instructions. 
% (3) We could need a more structured format for connection representation to be compatible with sequenced LM training loss to help model more easily learning the likelihood of certain connections or component selections.
% (4) Since the input of LM focuses on interpreting the values of voltage conversion ratio and efficiency, feeding these two numbers by text into LM could sacrifice the precision and let LM need additional efforts to understand the meaning of float numbers.

% \blue{need to include five-decimal statement.}

\subsection{Our Novel Circuit Formulations}

% In addition, this circuit representation cannot properly facilitate the graph learning with the language model training loss, i.e., the learning process cannot directly capture graph distance between logits and labels.
% As a result, we need a circuit representation that can ensure canonical form and help model to quantify the graph similarity through the learning process.

% \subsection{Instruction-based Data Format}  
% \begin{figure}[t]
% \centering
% \includegraphics[width=\linewidth]{fig/adjancency matrix representation.pdf}
% \vspace{-3mm}
% \caption{
% T5-styled masked language modeling~\cite{raffel2020exploring} for both edge and topology generation with our adjacency-matrix-based circuit formulation.
% }
% \label{fig:matrix_representation}
% \end{figure}

% Based on the identified potential shortcomings of the initial circuit format, our study proposes and systematically analyzes several alternative circuit representations to explore the most effective means of conveying complex circuit design specifications to an LM for automated analog circuit generation.

In addressing the potential shortcomings of the initial circuit format, our study introduces several alternative representations to more effectively convey complex circuit design specifications to an LM. 
In the meantime, we can explore the potential of LM for graph generation.

The first of these formulations, which we refer to as the Canonical Formulation, is illustrated in Figure~\ref{fig:different formulation}. 
By sorting the edges in connections based on a predefined vertex order, it organizes the circuit information to ensure a canonical, unambiguous representation for each design.
Furthermore, it simplifies the language used in the instructions, removing redundant or non-essential elements.  Thus, this method addresses the non-unique representation challenge.

Building on the Canonical Form, we introduced a second formulation, named Canonical Formulation + New Duty Cycle Representation, as shown in Figure~\ref{fig:different formulation}. This variant employs one-hot encoding for duty cycle selection, using 5 tokens of \textless select\textgreater \ or \textless unselect\textgreater.
Only one token is marked as \textless select\textgreater, indicating the chosen duty cycle.
This approach transforms a categorical choice into sequential selections.
% , thereby aligning with the sequential data processing nature of LMs. 
% Such alignment enhances the model's ability to accurately interpret and process the technical specifications inherent in circuit design.

% First formulation is "Duty cycle options 0.1 0.3 0.5 0.7 0.9, voltage conversion ratio 0.18887, efficiency 0.95544, Vertex order: VIN, VOUT, GND, Sa, Sa, Sb, Sb, L. Connections: (VIN, Sa0, L0), (VOUT, L1), (GND, Sb0), (Sb0, Sa1, L0), (L1, Sa0, Sa1) The duty cycle is set to 0.1."
% This will sort the edge in connections based on the vertex order to ensure there is a canonical form given an input for a specific circuit.

% Second formulation is "Duty cycle options 0.1 0.3 0.5 0.7 0.9, voltage conversion ratio 0.18887, efficiency 0.95544, Vertex order: VIN, VOUT, GND, Sa, Sa, Sb, Sb, L. Connections: (VIN, Sa0, L0), (VOUT, L1), (GND, Sb0), (Sb0, Sa1, L0), (L1, Sa0, Sa1) Duty cycle \textless select\textgreater \  \textless unselect\textgreater \  \textless unselect\textgreater \  \textless unselect\textgreater \  \textless unselect\textgreater"
% This explore the use of one-hot encoding to represent the duty cycle selection. 
% First \textless select\textgreater \ token means we select 0.1 as the duty cycle.

Building upon our second formulation, we present a third formulation, the Adjacency-Matrix-based Formulation, which offers a structured and systematic approach to represent circuit topologies. This method, depicted in Figure~\ref{fig:different formulation}, is designed to capture the graph structure effectively during training with an LM's loss function.

The formulation consists of two main parts: the prefix and the circuit description.
The prefix includes duty cycle options, voltage conversion ratio, and efficiency. 
The circuit description entails the duty cycle selection, vertex order, and connections.
% \red{The formulation consists of two main parts: the prefix, which includes duty cycle options, voltage conversion ratio, and efficiency, and the circuit description, which details the duty cycle selection, vertex order, and connections.
Also, this formulation separates each requirement with a \textless sep\textgreater\ token to aid the model in differentiating between them.

For the circuit description, duty cycle selection is succinctly represented using one-hot encoding from the second formulation.
Vertex order specifies the ports and devices in the circuit, while the connections are defined in an adjacency matrix format according to the vertex order.
Distinct tokens \textless no\_edge\textgreater, \textless edge\_1\textgreater, \textless edge\_2\textgreater, and \textless both\_edges\textgreater \ represent the presence or absence of connections between vertices. Note that \textless edge\_1\textgreater \ always precedes \textless edge\_2\textgreater \ in each vertex's connection representation. 
Thus, this formulation ensures a canonical form for each circuit.

By translating the graph structure into a sequential format, LMs can leverage their loss function to accurately predict the next item in the sequence, which in this case corresponds to the next connection or component in the circuit. The model can utilize the information from the preceding subgraph to learn the likelihood of connections, thereby understanding the graph structure more effectively.

Then, we introduce a fourth approach that innovatively feeds numerical data, including duty cycle options, voltage conversion ratio, and efficiency, as floats directly into the LM, which is called Float-input Adjacency-matrix-based Formulation. 
This needs a modification to the traditional word embedding mechanism of the LM. 
We incorporate a shared linear layer to encode these numbers, subsequently integrating them with other word embeddings to be the transformer's input.
% specifically designed to encode these numerical values before concatenating them with the word embeddings, preparing the input for the transformer architecture within the LM.

To apply matrix formulation for edge and topology generation, we employ T5-styled masked language modeling~\cite{raffel2020exploring}.
Each consecutive masked token will be replaced by one masking token.
Masking tokens indicate LM the positions where predictions are required.
For edge generation, two masking tokens are used to conceal the duty cycle selection and the connections, requiring the model to predict these aspects of the circuit based on the provided context. In topology generation, the scope of masking extends further to include vertex information, prompting the model to predict the entire circuit structure.
This method is pivotal in transitioning from edge to topology generation, allowing the model to incrementally build circuit understanding and refine its generative capabilities to mirror the iterative problem-solving nature of analog circuit.
% For edge generation, we use two masking tokens to conceal the duty cycle selection and connection. For topology generation, we further mask the vertex part.

% The forth formulation aims to feed the numbers in the prefix, including duty cycle options, voltage conversion ratio, and efficiency, with float to the LM.  
% We need to modify the mechanism of word embedding in LM for input the float number.
% Thus, we add a shared linear layer to encode the numbers in the prefix and then concatenate them with other word embedding for the input of transformer architecture.

For all formulations, we take the Flan-T5~\cite{chung2022scaling} as the pretrained model to perform SFT on our dataset. 
This is an encoder-decoder transformer structure with 248M parameters, which can facilitate our masked language modeling and handle the conditional generation for all formulations.
We initially train models on edge generation and then extend the training for topology generation to gradually advance the capabilities of LMs.
Furthermore, we integrate domain-specific tokens, such as VIN, Sa, and \textless edge\_1\textgreater, into the tokenizer. 
The expanded tokenizer ensures that these unique elements are recognized as single entities, improving the model ability to parse and generate the specialized content for circuit designs.
To enhance the robustness and generalization ability of our LM, we incorporate data augmentation techniques. 
Recognizing that the transformer backbone of the LM is not inherently permutation equivariant for graph, we introduce random vertex order permutations. 
This step exposes the model to diverse circuit configurations, enhancing its ability to generalize and preventing overfitting.

\section{Experimental Results}

\subsection{Experiment Setup}
% \red{Comparison of formulation based on T5 (autoregressive LM):
% naive formulation, naive formulation+cananical form, naive formulation+cananical form+remove redundent text, naive formulation+cananical form+remove redundent text+duty cycle,  pure text, float input text, pure float text}

% \red{Comparison of T5 with autoregressive LM and encoder-only LM}

% In evaluating the performance of our automated analog circuit design system, we conducted a series of experiments comparing some variants of circuit representation  to train the LLM. 

\noindent \textbf{Baselines.}
Since all related works~\cite{fan2021specification, zhao2022analog, lu2023automatic} focus on search-based algorithms, we are the first to construct a generative model to bridge specifications and circuits in a one-time generation approach.
Given the novelty of this application and the absence of prior work in building circuit generation models, our study aims to establish a baseline in the domain of automated analog circuit design, specifically focusing on the effectiveness of different input formulations. 
In addition, we comprehensively evaluate the model with 13K different input requirements from our testing set, while the other RL work~\cite{fan2021specification} only perform five different specifications to evaluate its search engine.

% \blue{First, we examine the effectiveness of few-shot prompt engineering using GPT4. Need experiemnts!!!!!!!!!}
We perform SFT on one baseline and four variants of circuit formulations for edge generation and topology generation in Figure~\ref{fig:different formulation}: (1) na\"ive formulation (NF), (2) our first formulation with canonical form (CF), (3) our second formulation with canonical form and one-hot-encoding-based duty cycle selection (CFDC), (4) our adjacency-matrix-based formulation with pure text input to LM (PM), and (5) our adjacency-matrix-based formulation with float input (FM). 
For NF, CF, CFDC, and PM, we set voltage conversion ratio and efficiency with a six-decimal precision in the input.
All models in the experiment were trained under identical hyperparameters, ensuring consistency across all other training variables. 
% \blue{Need to strictly state why not compare to baseline}
% \begin{enumerate}
%     \item Naive formulation
%     % \item Naive formulation + canonical form
%     \item Naive formulation + canonical form + removal of redundant text
%     \item Naive formulation + canonical form + removal of redundant text + duty cycle
%     \item Pure text adjacency-matrix-based formulation
%     \item Float input adjacency-matrix-based formulation
% \end{enumerate}

\noindent \textbf{Experimental platform and hyperparameters.}
Our experiment runs on a machine with one NVIDIA V100 GPU.
% with Intel\textsuperscript{\textregistered} Xeon\textsuperscript{\textregistered} E5-2687W CPUs.
% We apply the following training settings to construct our protector and realize the baseline methods:
The hyperparameters of the LM training are detailed as follows:
We perform training for 120 epochs using AdamW optimizer with a learning rate of $3 \times 10^{-4}$ with a cosine scheduler using 300 warmup steps, a batch size of $128$, and a L2 regularization strength of $10^{-5}$.

\noindent \textbf{Detailed model architecture.}
We use the encoder-decoder transformer structure with Flan-T5-base pretrained weights. It has 12 transformer layers in both encoder and decoder. Each layer has a key and value projection with dimension $64$, a feed-forward layer with dimension $2048$, and 12 heads.

\noindent \textbf{Evaluation metrics.} 
Our primary metrics for evaluation are the success rate of the generated circuits within varied tolerances and the Mean Squared Error (MSE) between the input specifications and the simulated performance of the generated circuits.

% \textbf{Metrics} 
The success rate, inspired by the code generation task~\cite{nijkamp2023codegen2}, is defined as the proportion of generated circuits whose simulated performance fell within a tolerance $t$ of the target input specifications. For instance, with a tolerance of $t$ = 0.1, a target input voltage and efficiency of 0.9 and 0.8, respectively, require the generated circuit's performance to be within the ranges of 0.9±0.1 for voltage and 0.8±0.1 for efficiency to be considered successful. 
In the experiment, we consider the success rate under 10 different $t$ ranging from 0.01 to 0.1.
MSE is computed as the average squared difference between the input specifications and the corresponding simulated performance metrics, providing a quantitative measure of the prediction accuracy.

\begin{figure*}[t]
\centering
\includegraphics[width=1\linewidth]{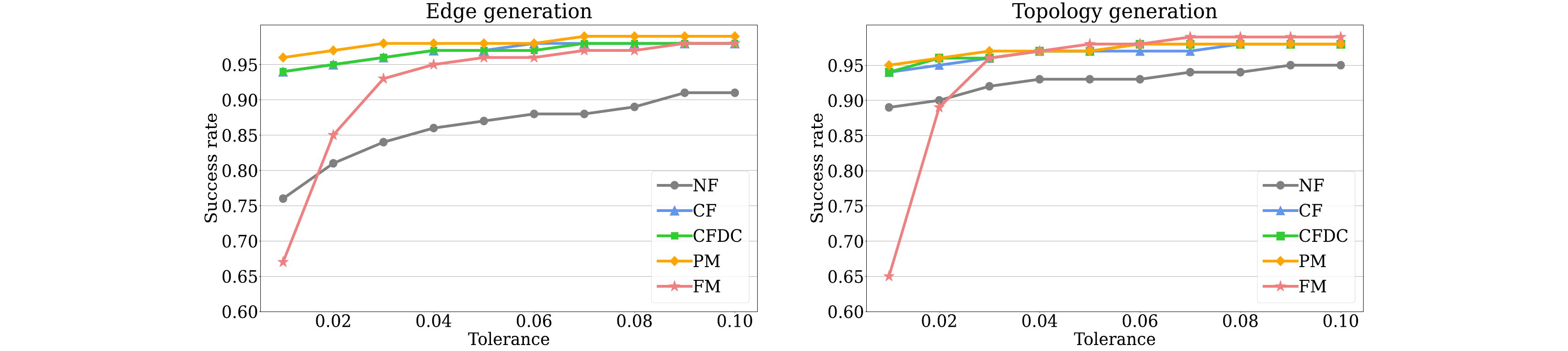}
\vspace{-3mm}
\caption{
Success rates of models trained with different circuit formulations using 3, 4, 5-component circuits for edge generation and topology generation task. Circuit formulations include (1) the na\"ive formulation (NF), (2) the canonical form (CF), (3) the canonical form with one-hot-encoding-based duty cycle selection (CFDC), (4) the adjacency-matrix-based formulation with pure text input (PM), and (5) the adjacency-matrix-based formulation with float input (FM).
}
\label{fig:success rate for 345-comp}
% \vspace{-3mm}
\end{figure*}

\begin{figure*}[t]
\centering
\includegraphics[width=1\linewidth]{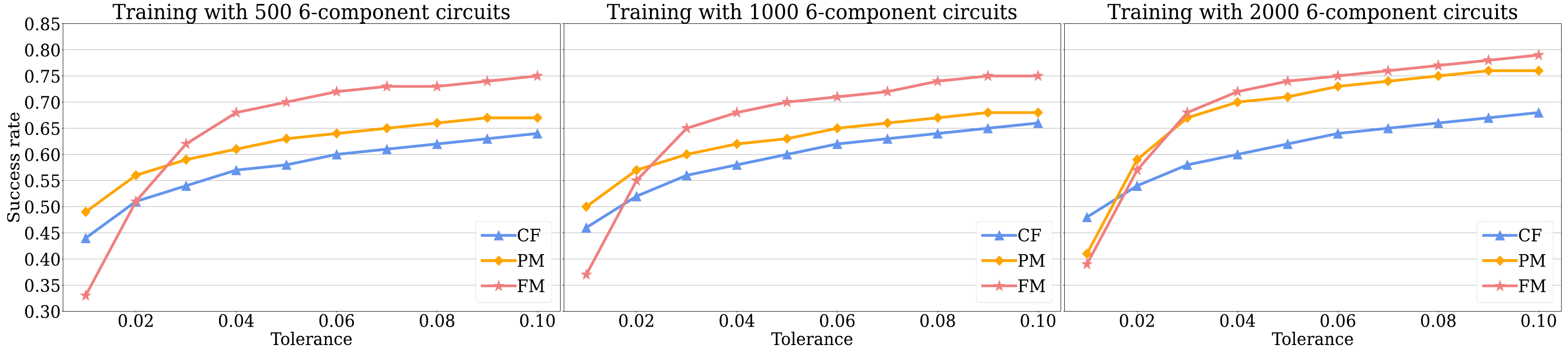}
\vspace{-3mm}
\caption{
Success rates of models finetuned with different circuit formulations using 500, 1000, and 2000 6-component circuits.
}
\label{fig:success rate for 6-comp}
% \vspace{-1mm}
\end{figure*}

\subsection{Generation Results on 3, 4, 5-Component Circuit} \label{sec:345 comp experiment}

% \begin{figure}[t]
% \centering
% \includegraphics[width=1\linewidth]{fig/success rate for edge generation.pdf}
% \vspace{-5mm}
% \caption{
% Success rates of models trained with different circuit formulations for edge generation task.
% }
% \label{fig:success rate edge}
% \end{figure}

% \begin{figure}[t]
% \centering
% \includegraphics[width=1\linewidth]{fig/success rate for topology generation.pdf}
% \vspace{-3mm}
% \caption{
% Success rates of models trained with different circuit formulations for topology generation task.
% }
% \label{fig:success rate topology}
% \vspace{-3mm}
% \end{figure}

The results of applying different circuit formulations for edge and topology generation are shown in Figure~\ref{fig:success rate for 345-comp} and Table~\ref{fig:MSE edge and topology}.
The proposed formulations, including CF, CFDC, PM, and FM, demonstrate a clear advantage over the baseline NF in both edge and topology generation tasks. This indicates the effectiveness of our formulations in guiding the model towards successful circuit generation.
\begin{table}[!t]\centering
\caption{MSE of voltage conversion ratio and efficiency evaluated on models trained with different circuit formulations for edge and topology generation. \vspace{1mm}}
\scriptsize
\begin{tabular}{c|cc|cc}\toprule
&\multicolumn{2}{c|}{Edge generation task} &\multicolumn{2}{c}{Topology generation task} \\\cmidrule{1-5}
MSE & Voltage & Efficiency & Voltage & Efficiency \\\cmidrule{1-5}
NF &0.054 &0.015 &0.031 &0.006 \\\cmidrule{1-5}
CF &0.016 &0.004 &0.008 &0.003 \\\cmidrule{1-5}
CFDC &0.007 &0.003 &0.005 &0.003 \\\cmidrule{1-5}
PM &0.007 & \textbf{0.002} &0.009 &0.003 \\\cmidrule{1-5}
FM &\textbf{0.006} &0.013 & \textbf{0.002} & \textbf{0.001} \\
\bottomrule
\end{tabular}
\label{fig:MSE edge and topology}
\vspace{-3mm}
\end{table}

For the edge generation task, while FM offered detailed numerical representation, it showed lower success rates at smaller tolerances compared to PM, CF, and CFDC. This can be attributed to FM's precise numerical approach lacking the inherent flexibility of text-based inputs, making exact value predictions more challenging within tight tolerance ranges. In contrast, PM, CF, and CFDC provide textual context, aiding the model in more quickly and accurately interpreting simpler circuit connections. 
In the topology generation task, FM's detailed numerical input shows advantageous at broader tolerances with success rates of 0.99, aligning with the complexity required in these scenarios. 

The discrepancy in FM's performance between edge and topology generation tasks suggests that numerical precision, while beneficial for complex tasks, may introduce complexities in tasks that require understanding simpler connections. The PM, CF, and CFDC, with their balance of structure and descriptive ease, lead to quicker and more accurate convergence in the edge generation task.

In addition, because encoder-only models theoretically offer the advantage of generating an entire graph in a single step, we experiment with using only the encoder component of T5, paired with PM, for circuit generation. 
Our result revealed that the encoder-only LM failed to converge and consistently produced invalid circuits, highlighting the importance of sequential reasoning and contextual understanding in autoregressive models for circuit design tasks.

In summary, our analysis not only affirms the necessity of task-specific circuit formulations but also emphasizes the suitability of autoregressive LMs over encoder-only LMs for analog circuit design. These findings are valuable for guiding future research directions and model selection in automated circuit generation.

\subsection{Transferability Evaluation on 6-Component Circuit} \label{sec:6 comp experiment}

\begin{table}[!tp]\centering
\caption{MSE of voltage conversion ratio and efficiency evaluated on models finetuned with different circuit formulations using 500, 1000, and 2000 6-component circuits.\vspace{1mm}}\label{tab: }
\scriptsize
\begin{tabular}{c|cc|cc|cc}\toprule
 &\multicolumn{2}{c|}{500} &\multicolumn{2}{c|}{1000} &\multicolumn{2}{c}{2000} \\\cmidrule{1-7}
MSE &Voltage &Efficency &Voltage &Efficency &Voltage &Efficency \\\cmidrule{1-7}
CF &0.109 &0.123 &0.097 &0.114 &0.082 &0.104 \\\cmidrule{1-7}
PM &0.092 &0.087 &0.096 &0.089 &0.051 & \textbf{0.053} \\\cmidrule{1-7}
FM &\textbf{0.050} & \textbf{0.068} & \textbf{0.048} & \textbf{0.049} & \textbf{0.038} & \textbf{0.053} \\
\bottomrule
\end{tabular}
\label{tab:mse for 6-comp}
\vspace{-3mm}
\end{table}

The scalability and adaptability of models in analog circuit design are crucial, particularly as the complexity of circuits increases. Our study initially focuses on models trained on circuits with 3, 4, and 5 components. We extend this to evaluate model performance on more complex 6-component circuits, providing essential insights into the circuit understanding capabilities of models in intricate scenarios.

We finetune LMs previously trained on 3, 4, 5-component circuits using three different formulations (CF, PM, and FM).
The finetuning is conducted on datasets of 500, 1000, and 2000 samples of 6-component circuits by given the model 6 component requirement for the edge generation task. 
Each larger dataset incorporates all circuits from the previous sets, allowing us to observe the performance trends as training data increased.
A testing set comprising 9k 6-component circuits is used for evaluation.

As demonstrated in Figure~\ref{fig:success rate for 6-comp} and Table~\ref{tab:mse for 6-comp}, FM particularly demonstrates superior adaptability and performance, especially at larger tolerances as the training data volume increased. 
PM also shows strong adaptability, with success rates higher than CF.
These performance gain can be attributed to the use of numerical inputs and the matrix formulation, which provide effectiveness for understanding complex circuit configurations.

We additionally train the model solely on 6-component circuits without pretraining on simpler circuits, and the model is unable to produce valid circuits. 
This finding underscores the importance of pretraining foundational models in analog circuit design, which is critical for understanding essential knowledge of basic circuit and effectively handling more advanced designs.

% \blue{emphasize the advantage of our methods, we don't need simulation, real application, how to use}

\begin{figure}[t]
\centering
\includegraphics[width=0.91\linewidth]{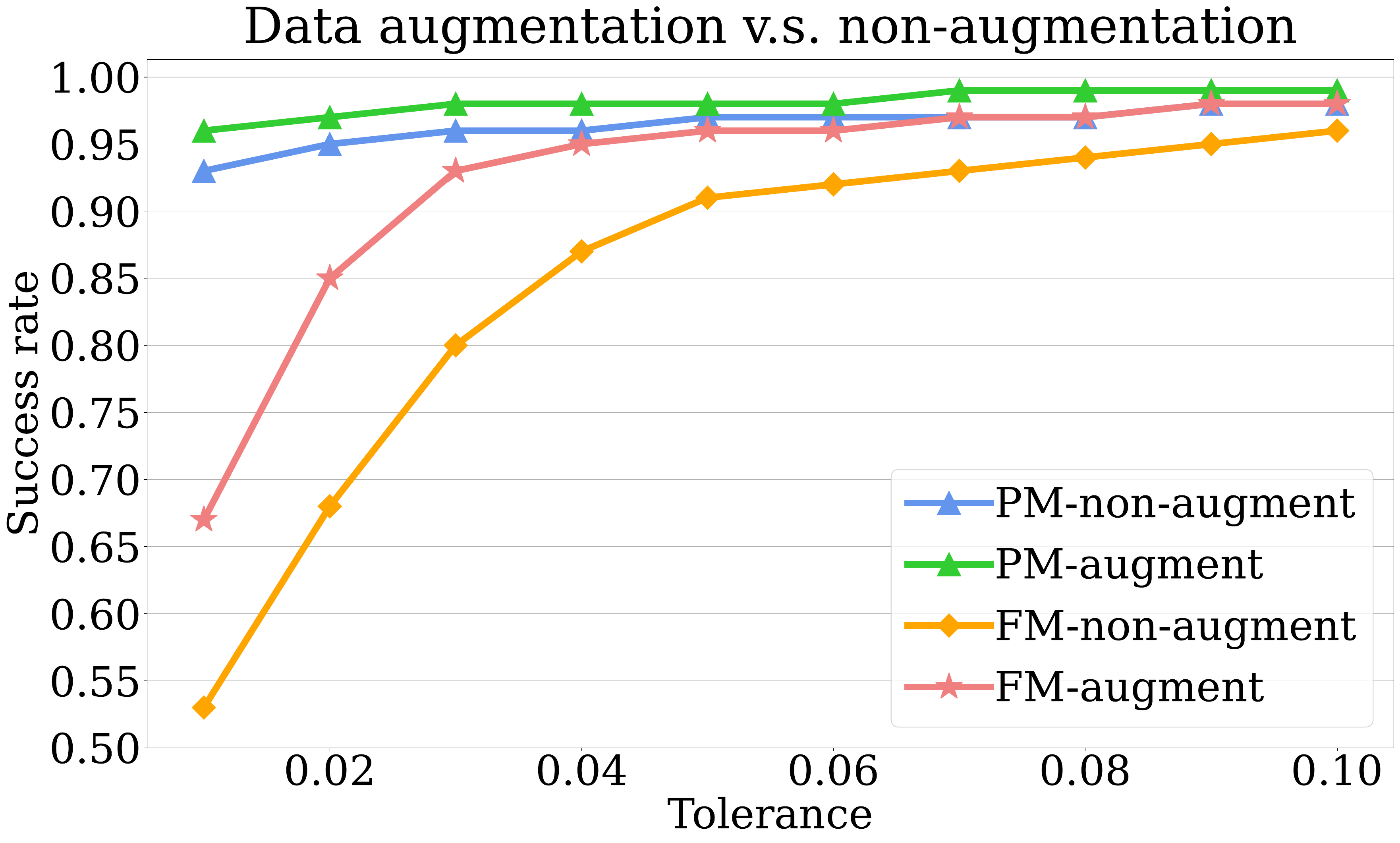}
% \vspace{-3mm}
\caption{
Success rates of models trained w/ and w/o vertex order permutation for edge generation.
}
\label{fig:success rate data augment}
% \vspace{-5mm}
\end{figure}

\begin{table}[!tp]\centering
\caption{MSE of voltage conversion ratio and efficiency evaluated on models trained w/ and w/o vertex order permutation for edge generation. \vspace{1mm}}\label{tab: mse data augment}
\scriptsize
\begin{tabular}{l|cc}
\toprule
&\multicolumn{2}{c}{Edge generation task} \\\cmidrule{1-3}
MSE &Voltage & Efficiency \\\cmidrule{1-3}
PM-non-augment &0.010 &0.005 \\\cmidrule{1-3} 
PM-augment & \textbf{0.007} & \textbf{0.002} \\\cmidrule{1-3} 
FM-non-augment &0.040 & \textbf{0.013} \\\cmidrule{1-3}
FM-augment & \textbf{0.006} & \textbf{0.013} \\
\bottomrule
\end{tabular}
% \vspace{-3mm}
\end{table}

\section{Discussion}

\subsection{Training with Vertex Order Permutation}

In this section, we evaluate the impact of data augmentation using random vertex order permutation in training that aims at enhancing the models' generation ability.
We train models using FM and PM with and without data augmentation for edge generation task, as shown in Figure~\ref{fig:success rate data augment} and Table~\ref{tab: mse data augment}.
The results indicate that models training with data augmentation outperform those training without data augmentation. 
These findings indicate the crucial role of data augmentation in increasing the diversity and complexity of training data. 
Additionally, this technique is particularly beneficial for FM, which initially struggle with lower success rates, to enhance the model to handle  numerical inputs.

\subsection{Exploration of Unseen Circuit Topology}

\begin{figure}[t]
\centering
\includegraphics[width=1\linewidth]{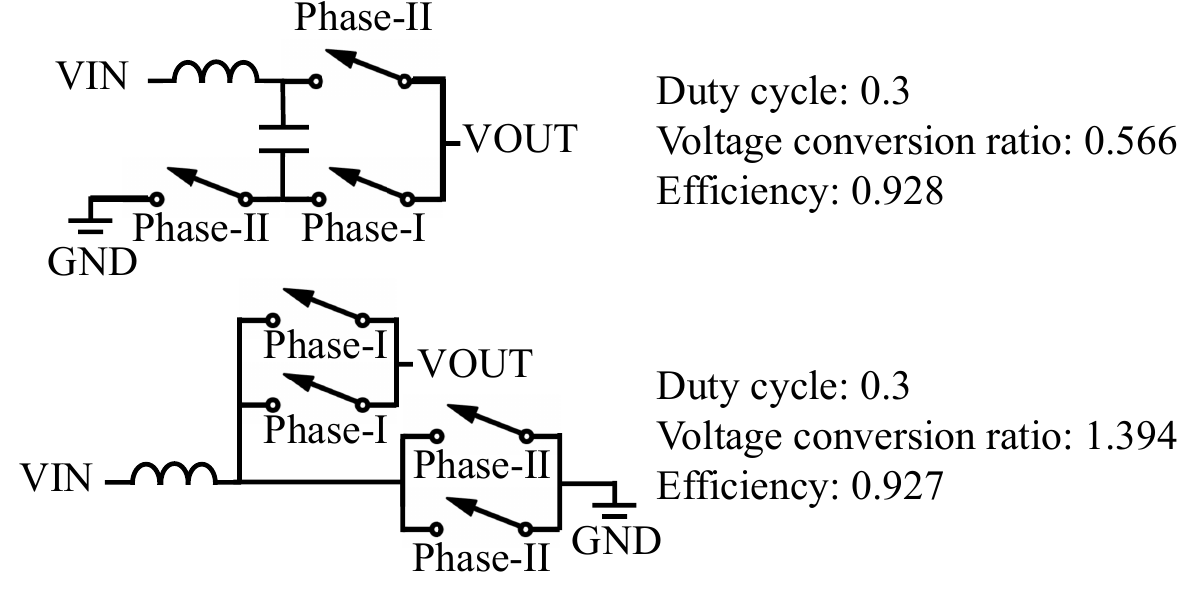}
% \vspace{-5mm}
\caption{Two promising power converter topologies generated by our LM, demonstrating its ability to design functional circuit configurations that satisfy target design specifications.
}
\label{fig:new circuit}
% \vspace{-3mm}
\end{figure}

\begin{table}[!t]\centering
\caption{Training vs. testing loss evaluated on models trained with different formulation for edge generation. \vspace{1mm}}
\label{tab:train_test_error}
\scriptsize
\begin{tabular}{l|cc} \toprule
&\multicolumn{2}{c}{Edge generation task} \\\cmidrule{1-3}
Loss & Training & Testing \\ \cmidrule{1-3}
CF   & 0.05  & 0.07 \\ \cmidrule{1-3}
CFDC & 0.05  & 0.07 \\ \cmidrule{1-3}
FM   & 0.07  & 0.09 \\ \cmidrule{1-3}
PM   & 0.03  & 0.04 \\ 
\bottomrule
\end{tabular}
% \vspace{-3mm}
\end{table}

In the field of automated analog circuit design, the ability to discover and validate previously unseen circuit topologies is important for newly developed tools. 
The majority of existing designs is created by human experts, and conventional automation tools typically lack the capability to generate unseen or novel designs.
According to experienced analog designers, a power converter with (1) a voltage conversion ratio ranging from 0.4 to 0.6 or greater than 1.2 and (2) an efficiency larger than 0.9 requires the most efforts to design manually.
Thus, we further validate that our LM can successfully generate high-performance power converter circuits within this target specification. 
Two promising circuits with 5 and 6 components are shown in Figure~\ref{fig:new circuit}. 
This achievement highlights the potential of our LM to serve as a valuable tool for designers to explore unseen design space for analog circuits, potentially pushing the boundaries of human designs in this area. 

% Efficiency 90\% up, voltage output 170, 180, or 40, 50

% \blue{Our experiment covers 10K input requirement, compare to previous work}

\subsection{Analysis of Potential Overfitting}
This section analyzes the potential for overfitting, a common challenge when employing sequential model architectures in graph generation tasks. The training and testing loss for our four different circuit formulations, CF, CFDC, PM, and FM, are presented in Table~\ref{tab:train_test_error}.

We observe more severe overfitting in CF and CFDC than FM and PM.
The robustness of FM and PM is further validated by their performance in transferability experiments detailed in Section~\ref{sec:6 comp experiment}, where these formulations achieve higher success rates in complex 6-component circuits. This evidence supports that FM and PM are more effective in avoiding overfitting and can be more suitable for scalable and generalizable circuit design tasks.

\iffalse
\subsection{Future Directions}
We propose three plans to manage larger circuits. As the dataset and model sizes used in our paper are relatively small compared to state-of-the-art LLMs, scaling up both model and dataset sizes is a primary step. We can leverage models pretrained on simpler circuits, progressively introducing them to more complex designs through finetuning, as shown in the experiment in Section 5.2. 

Second, to reduce the search space size, hierarchical decomposition can be used to segment circuit designs into smaller modules. These sub-tasks can be optimized individually before being integrated into large circuit designs. This mimics conventional human practice in complex analog design.

Third, we can investigate transformer decoding planning methods, which can do lookahead search to guide LMs towards generating better topologies. These approaches allow us to handle the expansive design space of larger circuits and will be included in future works.
\fi

% \subsection{Potential Overfitting}
% We examine the potential overfitting caused by the fact that we use sequential model architecture for graph generation modeling. We report the training and testing error in Table~\ref{tab:train_test_error} using our four formulations: CF, CFDC, PM, and FM. 
% We can observe more severe overfitting in CF and CFDC than FM and PM. This indicates that FM and PM are better formulations and can be further validated by the transferability experiment (Section~\ref{sec:6 comp experiment}), where FM and PM have higher success rates on 6-component circuits.
\section{Conclusion}
In this paper, we propose LaMAGIC, an LM-based topology generation model for analog circuit design that can directly generate an optimized circuit design given the custom specification in a single pass.
Our approach focuses on SFT and demonstrates the effectiveness of LMs in generating complex circuit topologies and deciding circuit parameters.
We systematically develop and analyze various input and output formulations to ensure canonical representations and the compatibility with the autoregressive nature of LMs, addressing the specific challenges of representing analog circuits as graphs.
Experimental results show that our novel circuit formulations can clearly outperform a naive formulation and achieve a success rate of up to 0.96 under a tolerance of 0.01. 
In addition, we examine the scalability and adaptability of our LMs on more complex six-component circuits. 
The results show that our proposed adjacency-matrix-based circuit formulation with float input to LM can have better effectiveness on complex circuit understanding, which can help the future research to focus on such formulations when dealing with complicated circuits.
% Overall, the methodologies and insights gained from this research demonstrate the potential of LMs in graph generation and offer a foundation for further exploration.
In the future, we will extend the capabilities of LaMAGIC to a wider range of analog components.
Through this article, we hope to open new research pathways on automated analog circuit design or any other fields of automated design that could benefit from the potential of LMs in graph generation.

\section*{Acknowledgements}
This work is partially supported by SRC 3104.001 and NSF 2106828.
Special thanks to Wan-Hsuan Lin for contributing valuable ideas to this project.

\section*{Impact Statement}
This paper presents work whose goal is to advance the field of Machine Learning. There are many potential societal consequences of our work, none which we feel must be specifically highlighted here.

% \begin{table}[t]
% \caption{Classification accuracies for naive Bayes and flexible
% Bayes on various data sets.}
% \label{sample-table}
% \vskip 0.15in
% \begin{center}
% \begin{small}
% \begin{sc}
% \begin{tabular}{lcccr}
% \toprule
% Data set & Naive & Flexible & Better? \\
% \midrule
% Breast    & 95.9$\pm$ 0.2& 96.7$\pm$ 0.2& $\surd$ \\
% Cleveland & 83.3$\pm$ 0.6& 80.0$\pm$ 0.6& $\times$\\
% Glass2    & 61.9$\pm$ 1.4& 83.8$\pm$ 0.7& $\surd$ \\
% Credit    & 74.8$\pm$ 0.5& 78.3$\pm$ 0.6&         \\
% Horse     & 73.3$\pm$ 0.9& 69.7$\pm$ 1.0& $\times$\\
% Meta      & 67.1$\pm$ 0.6& 76.5$\pm$ 0.5& $\surd$ \\
% Pima      & 75.1$\pm$ 0.6& 73.9$\pm$ 0.5&         \\
% Vehicle   & 44.9$\pm$ 0.6& 61.5$\pm$ 0.4& $\surd$ \\
% \bottomrule
% \end{tabular}
% \end{sc}
% \end{small}
% \end{center}
% \vskip -0.1in
% \end{table}

% Tables contain textual material, whereas figures contain graphical material.
% Specify the contents of each row and column in the table's topmost
% row. Again, you may float tables to a column's top or bottom, and set
% wide tables across both columns. Place two-column tables at the
% top or bottom of the page.
% x

\bibliographystyle{icml2024}

%%%%%%%%%%%%%%%%%%%%%%%%%%%%%%%%%%%%%%%%%%%%%%%%%%%%%%%%%%%%%%%%%%%%%%%%%%%%%%%
%%%%%%%%%%%%%%%%%%%%%%%%%%%%%%%%%%%%%%%%%%%%%%%%%%%%%%%%%%%%%%%%%%%%%%%%%%%%%%%
% APPENDIX
%%%%%%%%%%%%%%%%%%%%%%%%%%%%%%%%%%%%%%%%%%%%%%%%%%%%%%%%%%%%%%%%%%%%%%%%%%%%%%%
%%%%%%%%%%%%%%%%%%%%%%%%%%%%%%%%%%%%%%%%%%%%%%%%%%%%%%%%%%%%%%%%%%%%%%%%%%%%%%%
% \newpage
% \appendix
% \onecolumn

% You can have as much text here as you want. The main body must be at most $8$ pages long.
% For the final version, one more page can be added.
% If you want, you can use an appendix like this one.  

% The $\mathtt{\backslash onecolumn}$ command above can be kept in place if you prefer a one-column appendix, or can be removed if you prefer a two-column appendix.  Apart from this possible change, the style (font size, spacing, margins, page numbering, etc.) should be kept the same as the main body.
%%%%%%%%%%%%%%%%%%%%%%%%%%%%%%%%%%%%%%%%%%%%%%%%%%%%%%%%%%%%%%%%%%%%%%%%%%%%%%%
%%%%%%%%%%%%%%%%%%%%%%%%%%%%%%%%%%%%%%%%%%%%%%%%%%%%%%%%%%%%%%%%%%%%%%%%%%%%%%%

\end{document}